\documentclass[iop,revtex4]{emulateapj}
\usepackage{graphics,graphicx}
\usepackage{helvet}
\usepackage[utf8]{inputenc}
\usepackage[T1]{fontenc}
\usepackage{soul}
\usepackage{hyperref}
\usepackage{amsmath, amssymb, gensymb}
\usepackage{ulem}
\usepackage{natbib}
\usepackage{microtype}
\usepackage{multirow}
\usepackage{morefloats}
\usepackage{rotating}
\usepackage{xspace}
\usepackage{color}
\usepackage{xspace}
\def\kms {km\,s$^{-1}$\xspace}

\def\jybmkms {Jy\,beam$^{-1}$\,km\,s$^{-1}$\xspace}
\def\mjybmkms {mJy\,beam$^{-1}$\,km\,s$^{-1}$\xspace}
\def\jybm {Jy\,beam$^{-1}$\xspace}
\def\mjybm {mJy\,beam$^{-1}$\xspace}
\def\ujybm {$\mu$Jy\,beam$^{-1}$\xspace}
\def\mjybmvert {$\left(\frac{\textrm{mJy}}{\textrm{beam}}\right)$\xspace}

\def\jybmkms {Jy\,beam$^{-1}$\,km\,s$^{-1}$\xspace}

\def\vlsr {$v_{\textrm{LSR}}$\xspace}
\def\etal {\textit{et al.}\xspace}
\newcommand{\mir}[1]{\textbf{\fontfamily{lmvtt}\selectfont #1}} % font for code, e.g., MIRIAD

\def\sio{SiO\,($J$\,=\,5\,$\rightarrow\,$4)\xspace}

\def\co {CO\,($J$\,=\,2\,$\rightarrow\,$1)\xspace}

\def\cothreetotwo {CO\,($J$\,=\,3\,$\rightarrow\,$2)\xspace}
\def\cs {$^{13}$CS\,($J$\,=\,5\,$\rightarrow\,$4)\xspace}

\def\dco {DCO$^+$\,($J$\,=\,3\,$\rightarrow\,$2)\xspace}

\newcommand{\htcop}	{H$^{13}$CO$^+$\xspace}
\newcommand{\htcn}		{H$^{13}$CN\xspace}

\begin{document}

\slugcomment{Accepted for publication in ApJ on 12 July 2017}

\title{ALMA observations of dust polarization and molecular line emission from the Class 0 protostellar source Serpens SMM1}
\shorttitle{ALMA dust polarization \& molecular line observations of Serpens SMM1}

\author{Charles L. H. Hull\altaffilmark{1,2}}
\author{Josep M. Girart\altaffilmark{3}}
\author{\L{}ukasz Tychoniec\altaffilmark{4}}
\author{Ramprasad Rao\altaffilmark{5}}
\author{Paulo C. Cort\'es\altaffilmark{6,7}}
\author{Riwaj Pokhrel\altaffilmark{8,1}}
\author{Qizhou Zhang\altaffilmark{1}}
\author{Martin Houde\altaffilmark{9}}
\author{Michael M. Dunham\altaffilmark{10,1}}
\author{Lars~E.~Kristensen,\altaffilmark{11}}
\author{Shih-Ping Lai\altaffilmark{12,13}}
\author{Zhi-Yun Li\altaffilmark{14}}
\author{Richard L. Plambeck\altaffilmark{15}}

\altaffiltext{1}{Harvard-Smithsonian Center for Astrophysics, 60 Garden St., Cambridge, MA 02138, USA}
\altaffiltext{2}{Jansky Fellow of the National Radio Astronomy Observatory}
\altaffiltext{3}{Institut de Ci\`encies de l'Espai, (CSIC-IEEC), Campus UAB, Carrer de Can Magrans S/N, 08193 Cerdanyola del Vall\`es, Catalonia, Spain}
\altaffiltext{4}{Leiden Observatory, Leiden University, Niels Bohrweg 2, 2333 CA Leiden, The Netherlands}
\altaffiltext{5}{Institute of Astronomy and Astrophysics, Academia Sinica, 645 N. Aohoku Place, Hilo, HI 96720, USA}
\altaffiltext{6}{National Radio Astronomy Observatory, Charlottesville, VA 22903, USA}
\altaffiltext{7}{Joint ALMA Office, Alonso de C\'ordova 3107, Vitacura, Santiago, Chile}
\altaffiltext{8}{Department of Astronomy, University of Massachusetts, Amherst, MA 01003, USA}
\altaffiltext{9}{Department of Physics and Astronomy, The University of Western Ontario, London, ON N6A 3K7, Canada}
\altaffiltext{10}{Department of Physics, State University of New York at Fredonia, 280 Central Ave, Fredonia, NY 14063, USA}
\altaffiltext{11}{Centre for Star and Planet Formation, Niels Bohr Institute and Natural History Museum of Denmark, University of Copenhagen, \O{}ster Voldgade 5-7, DK-1350 Copenhagen K, Denmark} 
\altaffiltext{12}{Institute of Astronomy and Department of Physics, National Tsing Hua University, 101 Section 2 Kuang Fu Road, Hsinchu 30013,
Taiwan}
\altaffiltext{13}{Academia Sinica Institute of Astronomy and Astrophysics, P.O. Box 23-141, Taipei 10617, Taiwan}
\altaffiltext{14}{Department of Astronomy, University of Virginia, Charlottesville, VA 22903, USA}
\altaffiltext{15}{Astronomy Department \& Radio Astronomy Laboratory, University of California, Berkeley, CA 94720-3411, USA}

\shortauthors{Hull \etal}
\email{chat.hull@cfa.harvard.edu}

\begin{abstract}
We present high angular resolution dust polarization and molecular line observations carried out with the Atacama Large Millimeter/submillimeter Array (ALMA) toward the Class 0 protostar Serpens SMM1.  By complementing these observations with new polarization observations from the Submillimeter Array (SMA) and archival data from the Combined Array for Research in Millimeter-wave Astronomy (CARMA) and the James Clerk Maxwell Telescopes (JCMT), we can compare the magnetic field orientations at different spatial scales. We find major changes in the magnetic field orientation between large ($\sim$\,0.1\,pc) scales---where the magnetic field is oriented E--W, perpendicular to the major axis of the dusty filament where SMM1 is embedded---and the intermediate and small scales probed by CARMA ($\sim$\,1000\,AU resolution), the SMA ($\sim$\,350\,AU resolution), and ALMA ($\sim$\,140\,AU resolution).  The ALMA maps reveal that the redshifted lobe of the bipolar outflow is shaping the magnetic field in SMM1 on the southeast side of the source; however, on the northwestern side and elsewhere in the source, low velocity shocks may be causing the observed chaotic magnetic field pattern.  High-spatial-resolution continuum and spectral-line observations also reveal a tight ($\sim$\,130\,AU) protobinary system in SMM1-b, the eastern component of which is launching an extremely high-velocity, one-sided jet visible in both \co and \sio; however, that jet does not appear to be shaping the magnetic field.  These observations show that with the sensitivity and resolution of ALMA, we can now begin to understand the role that feedback (e.g., from protostellar outflows) plays in shaping the magnetic field in very young, star-forming sources like SMM1. 
\\ 
\end{abstract}


\section{Introduction}
\label{sec:intro}

The Serpens Main molecular cloud is an active star forming region, and the birthplace of a young cluster \citep[e.g.,][]{Eiroa2008}, located at a distance of 436\,$\pm$\,9\,pc \citep{OrtizLeon2017b}. The cloud is composed of a complex network of self-gravitating filaments where star formation is taking place \citep{Lee2014, Roccatagliata2015}; there is evidence that a cloud-cloud collision has triggered or enhanced the recent star formation in the region \citep{DuarteCabral2010, DuarteCabral2011}. 

Serpens SMM1,\footnote{Serpens SMM1 has been known by many names including Serpens FIRS1, Serp-FIR1, Ser-emb 6, IRAS 18273+0113, S68 FIR, S68 FIRS1, and S68-1b.} a Class 0 protostar, is the brightest millimeter source in the cloud \citep{Testi2000, Enoch2009, Lee2014}, with a luminosity $L_{\rm bol} = 100\,L_\odot$ \citep{Goicoechea2012}. It powers a compact ($\sim$\,2000\,AU), non-thermal radio jet that is expanding at velocities of $\sim$\,200\,\kms, which implies that the radio jet has a dynamical age of only 60\,yr \citep{Rodriguez1989, Curiel1993, Choi1999,RodriguezKamenetzky2016}; \citet{Curiel1993} suggest that the radio jet comprises a proto-Herbig-Haro system. The jet has a well collimated molecular outflow counterpart \citep{Curiel1996} that is also detectable in mid-infrared atomic lines \citep{Dionatos2010, Dionatos2014}; the jet appears to be perturbing the dense molecular gas surrounding the outflow cavity \citep{Torrelles1992}, producing copious water maser emission \citep{vanKempen2009}.  Atacama Large Millimeter/submillimeter Array (ALMA) observations from \citet{Hull2016a} show that the central source (SMM1-a; see Table \ref{table:sources}) powers an extremely high-velocity (EHV) molecular jet, which is surrounded by an ionized cavity detected in free-free emission by the VLA.  The cavity is most likely ionized either by the precessing high-velocity jet or by UV radiation from the central accreting protostar.

Polarized dust emission can be used as a tracer of magnetic fields in star-forming regions, as ``radiative torques'' \citep{Hoang2009} tend to align spinning dust grains with their long axes perpendicular to the ambient magnetic field \citep{Lazarian2007, Andersson2015}.  Dust polarization observations with (sub)millimeter interferometers have proven useful to trace the magnetic field at the dense core scales \citep[e.g.,][]{Rao1998, Girart1999, Lai2001, Alves2011, Hull2013, Hull2014}.   When a collapsing protostellar core is threaded by a uniform magnetic field and has low angular momentum \citep[relative to the magnetic energy;][]{Machida2005}, the magnetic field is expected to exhibit an hourglass morphology at the core scale, with the magnetic field orientation along the core's minor axis \citep{Fiedler1993, Galli1993b, Allen2003, Goncalves2008, Frau2011}. This morphology has been seen in some low- and high-mass protostars \citep{Lai2002, Girart2006, Girart2009, Rao2009, Tang2009b, Stephens2013, Qiu2014, HBLi2015}.  However, it is becoming clear that this situation is not universal: in several cases the magnetic fields threading the cores exhibit complex morphologies \citep[e.g.,][]{Tang2009a, Girart2013, Hull2014, Frau2014, Hull2017a}. In addition, recent observational studies of a large sample of star-forming sources \citep[][]{Hull2013, Hull2014} and analysis of synthetic observations of magnetohydrodynamic (MHD) simulations at similar resolution \citep{JLee2017} show no strong correlation between the outflow orientation and the core's magnetic field orientation at $\sim$\,1000\,AU scales,\footnote{The entire sample of observations from \citet{Hull2014} and the full suite of synthetic observations from \citet{JLee2017} showed random alignment of outflows with respect to magnetic fields.  However, weak correlations were found in subsets of the observations and simulations: in \citet{Hull2014}, the sources with low polarization fractions showed a slight tendency to have perpendicular outflows and magnetic fields; and in \citet{JLee2017}, the synthetic observations from the very strongly magnetized simulation showed a slight tendency to have aligned outflows and magnetic fields.} although there are studies that do suggest non-random alignment of outflows and magnetic fields at $\sim$\,10,000\,AU scales \citep[e.g.,][]{Chapman2013}.

In this paper we present ALMA 343\,GHz (Band 7) polarization observations toward the very embedded intermediate-mass protostar Serpens SMM1. We complement these observations with new Submillimeter Array (SMA; \citealt{Ho2004}) 345\,GHz dust polarization observations as well as with archival polarization maps obtained with the James Clerk Maxwell Telescope (JCMT) \citep{Davis2000, Matthews2009} and the Combined Array for Research in Millimeter-wave Astronomy (CARMA) \citep{Hull2014}.  The ALMA results we present here are among the first results from the ALMA full-polarization system, which has already led to publications on magnetized low- \citep{Hull2017a} and high-mass star formation \citep{Cortes2016}; quasar polarization \citep{Nagai2016}; and protostellar disk polarization \citep{Kataoka2016b}.  

In Section~\ref{sec:obs} we describe the observations and data reduction. In Section~\ref{sec:res} we present and describe the dust total intensity and polarization maps as well as the molecular line maps. In Section~\ref{sec:dis} we discuss the changes in magnetic field as a function of spatial scale and the relationship between the magnetic field and the outflows, jet, and dense-gas kinematics. Our conclusions are summarized in Section~\ref{sec:con}.

\begin{table}[hbt!]
\normalsize
\begin{center}
\caption{\normalsize \vspace{0.1in} SMM1 source properties}
\begin{tabular}{lccc}
\hline
\hline
Name & $\alpha_{\textrm{J2000}}$ & $\delta_{\textrm{J2000}}$ & $I_{870}$ \\
 & & & \small (\mjybm) \\
\hline
SMM1-a & 18:29:49.81 & +1:15:20.41 & 800  \\ 
SMM1-b & 18:29:49.67 & +1:15:21.15  & 106  \\ 
SMM1-c & 18:29:49.93 & +1:15:22.02 & 28.1  \\ 
SMM1-d & 18:29:49.99 & +1:15:22.97 & 10.1  \\
\hline
\end{tabular}
\label{table:sources}
\end{center}
\footnotesize
\textbf{Note.} Properties of the four continuum sources detected in the ALMA data (Figure \ref{fig:pol}(d), grayscale).  $I_{870}$ is the peak intensity of each of the sources in the 870\,$\micron$ ALMA data.
\end{table}

\section{Observations}
\label{sec:obs}

\begin{figure*}[]
\vspace{0.4in}
\includegraphics[width=\linewidth, clip, trim=0cm 0cm 0cm 0cm]{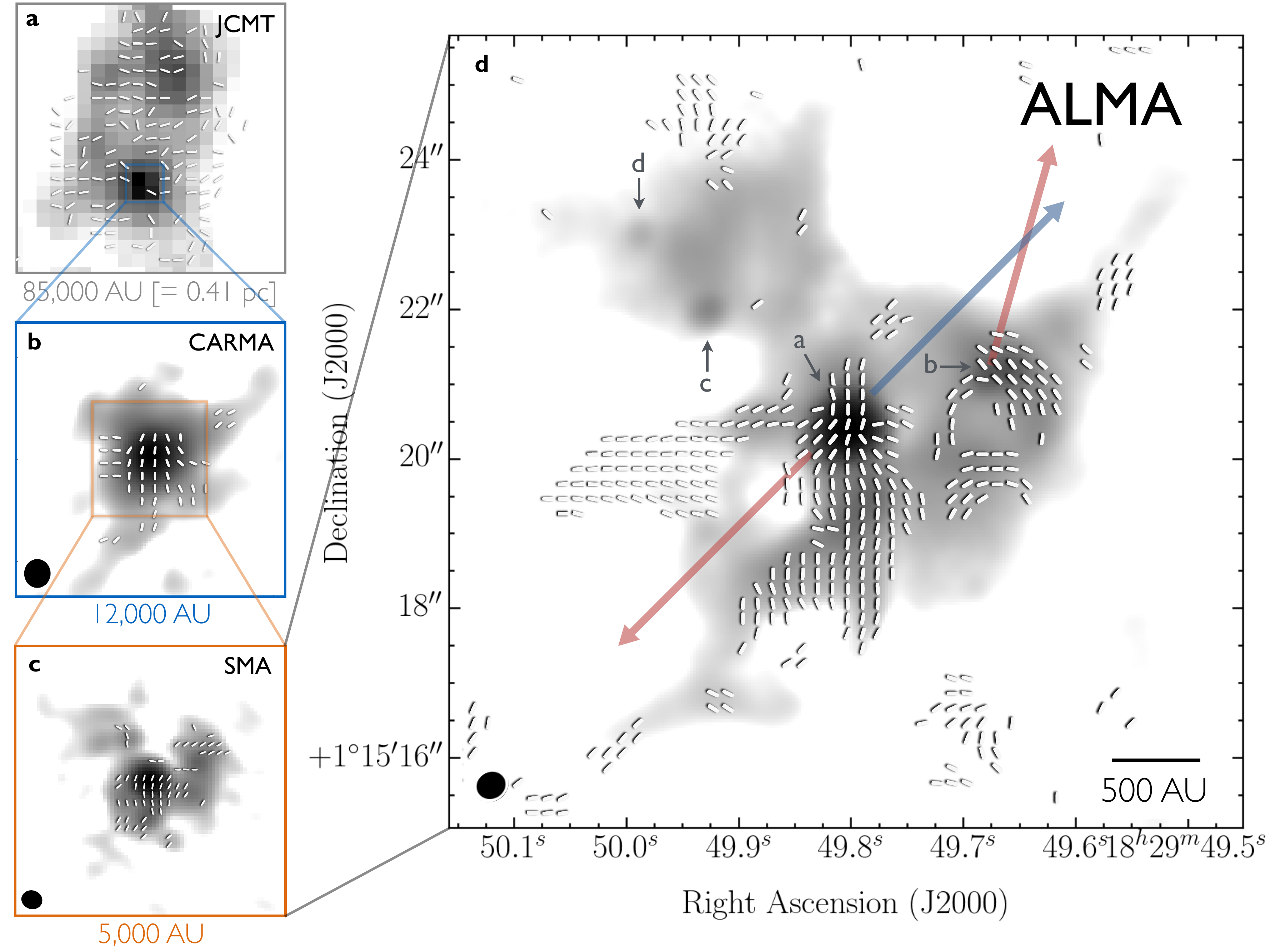}
\vspace{0.1in}
\caption{\footnotesize 
Multi-scale view of the magnetic field around Serpens SMM1. 
Line segments represent the magnetic field orientation, rotated by 90$\degree$ from the dust polarization (the length of the line segments in each panel is identical, and does not represent any other quantity).
Grayscale is total intensity (Stokes $I$) thermal dust emission.  
Panel (a) shows 850\,\micron{} JCMT observations \citep{Matthews2009}, (b) shows 1.3\,mm CARMA observations \citep{Hull2014}, (c) shows 880\,\micron{} SMA observations, and (d) shows 870\,\micron{} ALMA observations.
For the 880\,\micron{} SMA data, line segments are plotted where the polarized intensity $P > 2\,\sigma_P$; the rms noise in the polarized intensity map $\sigma_P = 2$\,\mjybm.  The dust emission is shown starting at 2 $\times$ $\sigma_I$, where the rms noise in the Stokes $I$ map $\sigma_I$ = 4\,\mjybm.  The peak total intensity in the SMA data is 1.43\,\jybm.
For the 870\,\micron{} ALMA data, line segments are plotted where the polarized intensity $P > 3\,\sigma_P$; the rms noise in the polarized intensity map $\sigma_P = 60$\,\ujybm.  The dust emission is shown starting at 3 $\times$ $\sigma_I$, where the rms noise in the Stokes $I$ map $\sigma_I$ = 0.5\,\mjybm.  The peak polarized and total intensities in the ALMA data are 11.8\,\mjybm and 800\,\mjybm, respectively.  
The red and blue arrows originating at the central source (SMM1-a) are the red- and blue-shifted lobes of the bipolar outflow from SMM1-a traced in \co (see Figure \ref{fig:alma_pol_co}).  The red arrow originating at SMM1-b (the source to the west of SMM1-a) is the redshifted EHV \sio jet shown in Figure \ref{fig:sio}.
The text below each of the panels on the left indicates the physical size of the image at the 436\,pc distance to the Serpens Main region.  
The black ellipses in the lower-left corners of the ALMA, SMA, and CARMA maps represent the synthesized beams (resolution elements).  The ALMA beam measures $0\farcs35\times0\farcs32$ (146\,AU at a distance of 436\,pc) at a position angle of --61$\degree$; the SMA beam measures $0\farcs86\times0\farcs75$ (350\,AU) with a position angle of 74$\degree$; and the CARMA beam data measures $2\farcs90\times2\farcs46$ (1165\,AU) at a position angle of 9$\degree$.  The JCMT data have a resolution of $20\arcsec$ (8720\,AU). 
Each of the four sources (SMM1-a, b, c, and d) are indicated in panel (d); source properties can be found in Table \ref{table:sources}.
The details of all four datasets are summarized in Table \ref{table:obs}.
\textit{The ALMA data used to make the figure in panel (d) are available in the online version of this publication.}
}
\label{fig:pol}
\end{figure*}

\subsection{ALMA observations}

The 870\,\micron{} ALMA dust polarization observations that we present were taken on 2015 June 3 and 2015 June 7, and have a synthesized beam (resolution element) of $\sim$\,$0\farcs33$, corresponding to a linear resolution of $\sim$\,140\,AU at a distance of 436\,pc.  The largest recoverable scale in the data is approximately 5$\arcsec$.  The ALMA polarization data comprise 8\,GHz of wide-band dust continuum ranging in frequency from $\sim$\,336--350\,GHz, with a mean frequency of 343.479\,GHz (873\,$\micron$).  
The main calibration sources such as bandpass, flux, and phase are selected at run time by querying the ALMA source catalog. The polarization calibrator was selected by hand to be J1751+0939 because of its high polarization fraction. This source was also selected by the online system as the bandpass and phase calibrator.  Titan was selected as the flux calibrator. The ALMA flux accuracy in Band 6 (1.3\,mm) and Band 7 (870\,\micron{}) is $\sim$\,10\%, as determined by the observatory flux monitoring program. The gain calibration uncertainty is $\sim$\,5\% in Band 6 and $\sim$\,10\% in Band 7. The accuracy in the bandpass calibration is $\lesssim$\,0.2\% in amplitude and $\lesssim$\,0.5$\degree$ in phase.  For a detailed discussion of the ALMA polarization system, see \citet{Nagai2016}.

The dust continuum image, most clearly seen in Figure \ref{fig:pol}(d), was produced by using the CASA task \mir{CLEAN} with a Briggs weighting parameter of robust\,=\,1.  The image was improved iteratively by four rounds of phase-only self-calibration using the total intensity (Stokes $I$) image as a model.  The Stokes $I$, $Q$, and $U$ maps (where the $Q$ and $U$ maps show the polarized emission) were each \mir{CLEANed} independently with an appropriate number of \mir{CLEAN} iterations after the final round of self-calibration.  The rms noise level in the final Stokes $I$ dust map is $\sigma_I = 0.5$\,\mjybm{}, whereas the rms noise level in the Stokes $Q$ and $U$ dust maps is $\sigma_Q \approx \sigma_U \approx \sigma_P = 0.06$\,\mjybm{}, where $\sigma_P$ is the rms noise in the map of polarized intensity $P$ (see Equation \ref{eqn:P} below).  The reason for this difference is that the total intensity image is more dynamic-range limited than the polarized intensity images.  This difference in noise levels allows one to detect polarized emission in some regions where one cannot reliably detect continuum dust emission.

The quantities that can be derived from the polarization maps are the polarized intensity $P$, the fractional polarization $P_\textrm{frac}$, and the polarization position angle $\chi$:

\begin{align}
P &= \sqrt{Q^2 + U^2} \label{eqn:P} \\
P_\textrm{frac} &= \frac{P}{I} \\
\chi &= \frac{1}{2} \arctan{\left(\frac{U}{Q}\right)}\, .
\end{align}

\noindent
Note that $P$ has a positive bias because it is always a positive quantity, even though the Stokes parameters $Q$ and $U$ from which $P$ is derived can be either positive or negative.  This bias has a particularly significant effect in low-signal-to-noise measurements.  We thus debias the polarized intensity map as described in \citet{Vaillancourt2006} and \citet{Hull2015b}.  See Table \ref{table:data} for the ALMA polarization data.

% Frequencies: mean([336494.5750, 338432.0750, 348494.5750, 350494.5750])

We also present 1.3\,mm (Band 6) ALMA spectral line data, which were taken in two different array configurations on 2014 August 18 ($\sim$\,0.3$\arcsec$ angular resolution) and 2015 April 06 ($\sim$\,1$\arcsec$ resolution).  These data include dust continuum as well as \co, which we use to image the outflow from SMM1 (see Figure \ref{fig:alma_pol_co} and \citealt{Hull2016a}); \sio (Figure \ref{fig:sio}); and \dco (Figure \ref{fig:dco+}).

Finally, we present 1.3\,mm ALMA continuum data with $\sim$\,$0\farcs1$ resolution (Pokhrel et al., in prep.), observed on 2016 Sep 10, 2016 Sep 13, and 2016 Oct 31.  These data show that SMM1-b is a binary with $\sim$\,130\,AU separation, and which we use to pinpoint the driving source of the high-velocity SiO jet (see Section \ref{sec:MolEm} and Figure \ref{fig:sio}).

\subsection{SMA observations}

The SMA polarization observations (Figure \ref{fig:pol}(c)) were taken on 2012 May 25 (compact configuration) and 2012 September 2 and 3 (extended configuration), and have a synthesized beam of $\sim$\,$0\farcs8$.
%The phase center of the telescope was $\alpha$(J2000.0)$=18^{\rm h}29^{\rm m}49\fs79$ and  $\delta$(J2000.0)$= 1\degr15\arcmin 20\farcs40$.  
In the May observations the frequency ranges covered were 332.0--336.0\,GHz and 344.0--348.0\,GHz in the lower sideband (LSB) and upper sideband (USB), respectively.  The ranges were slightly different for the September observations: 332.7--336.7\,GHz (LSB) and 344.7--348.7\,GHz (USB).   The correlator provided a spectral resolution of about 0.8\,MHz, or 0.7\,\kms at 345\,GHz.   The gain calibrator was the quasar J1751$+$096. The bandpass calibrator was BL Lac. The absolute flux scale was determined from observations of Titan.  The flux uncertainty was estimated to be $\sim$\,20\%.  The data were reduced using the software packages \mir{MIR} (see \citealt{QiYoung2015} for a description of how to reduce full-polarization data in \mir{MIR}) and \mir{MIRIAD} \citep{Sault1995}.

The SMA conducts polarimetric observations by cross correlating orthogonal circular polarizations (CP). The CP is produced by inserting quarter wave plates in front of the receivers, which have native linear polarization. The instrumentation techniques and calibration issues are discussed in detail in \citet{MarroneThesis} and \citet{Marrone2008b}.  The instrumental polarization (``leakage'') calibrator was chosen to be BL Lac, which was observed over a parallactic angle range of $\sim$\,60$\arcdeg$. We found polarization leakages between 1--2\% for the USB, while the LSB leakages were between 2--4\%. These leakages were measured to an accuracy of 0.1\%. 

We performed self-calibration using the continuum data and applied the derived gain solutions to the molecular line data.  We produced maps with natural weighting (robust\,=\,2) after subtracting the dust continuum emission in the visibility space. Table~\ref{table:T1} in Appendix \ref{appendix:lines} gives the transitions, frequencies, and lower energy levels of the molecular lines detected.

\subsection{JCMT and CARMA observations}

The archival JCMT SCUBA polarization data (Figure \ref{fig:pol}(a)) were obtained from supplementary data provided by \citet{Matthews2009}. These data were first published by \citet{Davis2000}; \citet{Matthews2009} performed a fresh reduction of the original \citet{Davis2000} data with a resulting angular resolution of  $\sim$\,20$\arcsec$. 

The CARMA polarization data (Figure \ref{fig:pol}(b)) were taken between 2011 and 2013 as part of the TADPOL survey \citep{Hull2014}, the largest high-resolution ($\sim$1000 AU) interferometric survey to date of dust polarization in low-mass star-forming cores.  The data were taken using the 1.3 mm polarization receiver system in the C, D, and E arrays at CARMA, which correspond to angular resolutions at 1.3\,mm of approximately 1$\arcsec$, 2$\arcsec$, and $4\arcsec$, respectively.  The details of the CARMA polarization system can be found in \citet{Hull2015b}; for descriptions of the observational setup and the data reduction procedure, see Section 3 of \citet{Hull2014}.  The image of the CARMA data in Figure~\ref{fig:pol} is an improved version of Figure 27 in \citet{Hull2014}, as the data presented here have been self-calibrated using the Stokes $I$ \mir{CLEAN} components as a model.

\begin{table*}[hbt!]
\normalsize
\begin{center}
\caption{\normalsize \vspace{0.1in} Observational details}
\begin{tabular}{cccccc}
\hline
\hline
Telescope & $\lambda$ & $\theta_{\textrm{res}}$ & $\theta_{\textrm{MRS}}$ &  $I_{\textrm{peak}}$ & $I_{\textrm{rms}}$ \\
                 &                   &                                      &              ($\arcsec$)         & \small (\jybm) & \small (\mjybm) \\
\hline
ALMA    & 870\,$\micron$   & $0\farcs35\times0\farcs32$  & 5.2   & 0.80 & 0.5 \\ 
% MRS: from Band 7 MRS value in ALMA Cycle 2 docs
SMA      & 880\,$\micron$   & $0\farcs86\times0\farcs75$ & 14.5  & 1.43 & 4.0 \\ 
% MRS calculation: 1.22 * 880 micron / 15.3 m radians in arcseconds, based on min baseline of compact config
CARMA & 1.3\,mm             & $2\farcs90\times2\farcs46$  & 41     & 1.30 & 6.2 \\ 
% MRS calculation: 1.22 * 1.3 mm / 8 m radians in arcseconds, based on min baseline of E config
JCMT$^a$    & 850\,$\micron$  & $20\arcsec$                          & ---     & 4.00 & --- \\
\hline
\end{tabular}
\label{table:obs}
\end{center}
\footnotesize
\textit{Note:} $\lambda$ is the wavelength of the observations.  $\theta_{\textrm{res}}$ is resolution of the observations, which, in the case of ALMA, the SMA, and CARMA, is the same as the synthesized beam of the interferometric data.  $\theta_{\textrm{MRS}}$ is the maximum recoverable scale in the interferometric data, calculated using the shortest baseline in each observation.  $I_{\textrm{peak}}$ and $I_{\textrm{rms}}$ are the peak total intensity and the rms noise in the total intensity maps, respectively; the values are calculated as flux density per synthesized beam $\theta_{\textrm{res}}$.

\smallskip
$^a$ For a discussion of the single-dish JCMT observations, noise estimates, and peak fluxes, see \citet{Matthews2009} (including Figure 56).
\vspace{1em}
\end{table*}

\section{Results}
\label{sec:res}

Below we discuss in detail a number of results from our continuum and spectral line observations of Serpens SMM1.  We begin by describing Figure \ref{fig:pol}, which shows the total-intensity and polarized dust emission toward SMM1 at various spatial scales using observations from the JCMT, CARMA, the SMA, and ALMA.  We then present molecular emission maps from ALMA, including \co (Figure \ref{fig:alma_pol_co}), which shows how the outflow is shaping the magnetic field; high-velocity \sio (Figure \ref{fig:sio}, right panel), which reveals an EHV jet emanating from SMM1-b; and \dco (Figure \ref{fig:dco+}) and low-velocity \sio (Figure \ref{fig:sio}, left panel), which trace the dense gas in which the protostars are embedded.

\subsection{Total-intensity and polarized dust emission}

Here we present the magnetic field derived from the polarized dust emission at the different scales as traced by different telescopes, moving from large to small scales. 

\textit{JCMT data:} The JCMT 850\,$\micron$ dust polarization map (Figure \ref{fig:pol}(a)) covers the whole $\sim 0.4$\,pc molecular clump where the SMM1 and SMM9\footnote{\,SMM9 is also known as S68N and Ser-emb~8; see \citealt{Hull2017a}.} dense cores are embedded.  \citet{Davis2000} found that the magnetic field is relatively uniform and is approximately perpendicular to the major axis of this clump, oriented E--W with a mean position angle of $\sim$\,80$\degree$. These authors found a magnetic field strength of $\sim 1$\,mG, estimated using the Davis-Chandrasekhar-Fermi (DCF) technique \citep{Davis1951,Chandrasekhar1953}.\footnote{If we take into account the calibration correction to the DCF technique developed by \citet{Ostriker2001}, the expected strength would be a factor of two lower, or $\sim$\,0.5\,mG \citep[see also][]{Falceta2008}.}

While the magnetic field is well ordered in the E--W direction, there is strong depolarization toward the emission peak of SMM1.  This is the ``polarization hole'' phenomenon, where the polarization fraction drops near the dust emission peak.  This phenomenon appears in both high- and low-resolution observations of star-forming cores \citep{Dotson1996, Girart2006, Liu2013} and simulations \citep{Padoan2001, Lazarian2007, Pelkonen2009, JLee2017}.  One possible cause of the polarization hole is that the plane-of-sky magnetic field could have structure on <\,20$\arcsec$ scales that cannot be resolved by the JCMT; this plane-of-sky averaging would reduce the polarization fraction.  And indeed, as we zoom into smaller scales in Figure \ref{fig:pol} we see more and more complicated magnetic field morphology in the higher resolution CARMA, SMA, and ALMA maps. 

\textit{CARMA data:} Figure \ref{fig:pol}(b) shows the 1.3\,mm  dust emission and the magnetic field derived from CARMA, with a resolution of $\sim$\,2$\farcs$5.  These are interferometric observations, and thus they are not sensitive to structures $\gtrsim\,15\arcsec$ (or $\sim$\,6000\,AU) in extent.  The magnetic field in the center of SMM1, undetected with the JCMT, is revealed by CARMA to be significantly different from the overall E--W orientation seen in the JCMT data: in the interferometric data, the field near the center of the source appears to be oriented predominantly in the N--S direction.  

\textit{SMA data:} As a comparison, Figure \ref{fig:pol}(c) shows the 880\,$\micron$ SMA map, which has an even higher resolution of $\sim$\,$0\farcs8$.  The magnetic field derived from the SMA and CARMA data are consistent toward the peak of SMM1.  Away from the dust emission peak, both the SMA and the CARMA data show hints that some regions of the magnetic field are oriented along the outflow, consistent with what is seen in the ALMA data (see Figure \ref{fig:alma_pol_co}).  Note that the E--W magnetic field component detected to the east of the source peak in both the CARMA and the ALMA data is not detected by the SMA, most likely due to a combination of dynamic range, signal-to-noise, and the scales recoverable from the higher resolution SMA data.

\textit{ALMA data:} Finally, we arrive at the 870\,$\micron$ ALMA map, which can be seen in Figure \ref{fig:pol}(d), and which achieves a resolution of $\sim$\,$0\farcs33$, or $\sim$\,140\,AU.  There are two main sources detected in the ALMA maps. Following \citet{Choi2009, Dionatos2014, Hull2016a}, we will refer to the brighter eastern source as SMM1-a and the fainter source $\sim$\,2$\arcsec$ to the WNW as SMM1-b. There are two compact but weaker sources northeast of SMM1-b, which we deem SMM1-c and SMM1-d. SMM1-c has a 3.6\,cm counterpart \citep[see Figure 1 from][]{Hull2016a}; such long-wavelength emission cannot be from dust, but rather is tracing ionized gas, suggesting that this source is an embedded protostellar object. SMM1-d has no known counterpart at other wavelengths, although it appears to be the source driving a low-velocity \sio outflow (see Section \ref{sec:MolEm} and Figure \ref{fig:sio}).  Coordinates and peak intensities of all four of the aforementioned sources are listed in Table \ref{table:sources}, and each source is indicated in Figure \ref{fig:pol}(d).

It is immediately apparent that the N--S magnetic field orientation that dominates the center of the CARMA and SMA maps is due to the bright, highly polarized emission extending southward from the peak of SMM1-a.  However, the ALMA data also show a very clear E--W feature in the magnetic field extending to the east of SMM1-a; both the N--S and E--W features are clearly tracing the edge of the low-velocity bipolar outflow pictured in Figure \ref{fig:alma_pol_co}.  The E--W feature can be seen in the CARMA map (Figure \ref{fig:pol}(b): see the few E--W line segments to the east of the SMM1-a peak), but at a much lower signal-to-noise than the N--S feature that otherwise dominates the lower resolution CARMA and SMA maps because of its much brighter polarized emission (see Section \ref{sec:pol_bias} for a discussion of this issue).  However, to the west of SMM1-a, the magnetic field does not have a preferred orientation and appears relatively chaotic. Indeed, around SMM1-b the magnetic field direction is neither parallel nor perpendicular to the fast, highly collimated jet associated with this source (see Figure \ref{fig:sio}). Northeast of SMM1-a, around SMM1-c and SMM1-d, there is very little polarization detected; dividing the rms noise level in this region by the detected Stokes $I$ intensity yields upper limits on the polarization fraction as low as a few $\times$ 0.1\%.

\subsection{Molecular emission}
\label{sec:MolEm}

\begin{figure} [hbt!]
\centering
\includegraphics[scale=0.3, clip, trim=0cm 0cm 0cm 0cm]{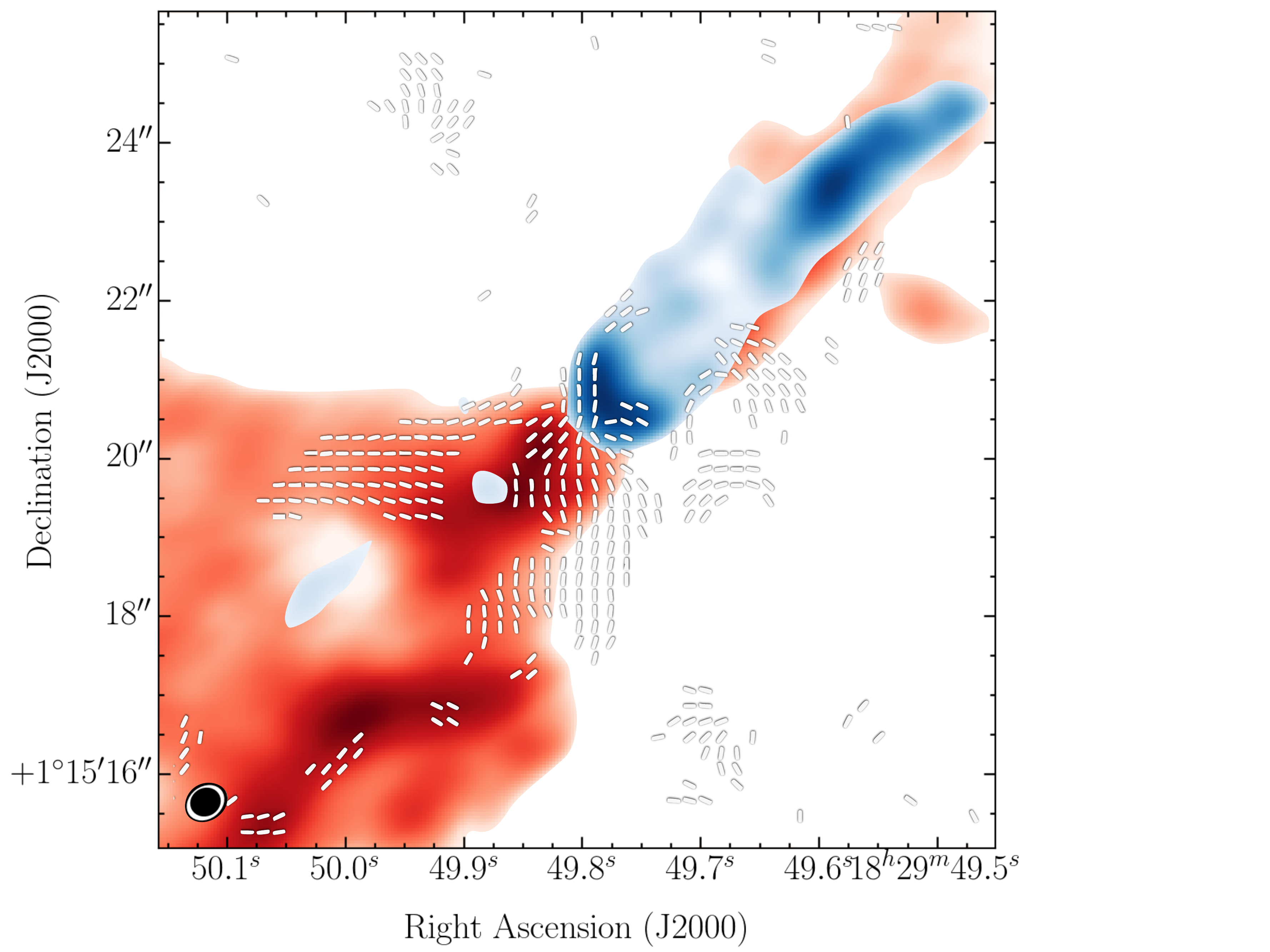}
\caption{\footnotesize   
Low-velocity red- and blueshifted \co from the ALMA data (red and blue color scales, respectively), adapted from \citet{Hull2016a}.  The CO velocity ranges are 2~to~15\,\kms (redshifted) and --20~to~--5\,\kms (blueshifted) relative to the \vlsr of SMM1 of $\sim$\,8.5\,\kms \citep{Lee2014}.  
The peaks of the redshifted and blueshifted moment 0 maps are 3.76 and 4.16\,\jybmkms{}, respectively.  
Line segments represent the inferred magnetic field orientation, reproduced from Figure \ref{fig:pol}(d).
The solid ellipse indicates the synthesized beam of the ALMA dust polarization data (see Figure \ref{fig:pol}); the larger open ellipse is the beam of the \co data, which measures $0\farcs55\times0\farcs45$ at a position angle of --53$\degree$.
% ORIGINAL VLSR VALUES: 12 to 25 km/s (redshifted) and --10 to 5 km/s (blueshifted).
} 
\vspace{0.2in}
\label{fig:alma_pol_co} 
\end{figure}

\begin{figure*} [hbt!]
\centering
\includegraphics[width=\linewidth, clip, trim=0cm 8.5cm 0cm 0cm]{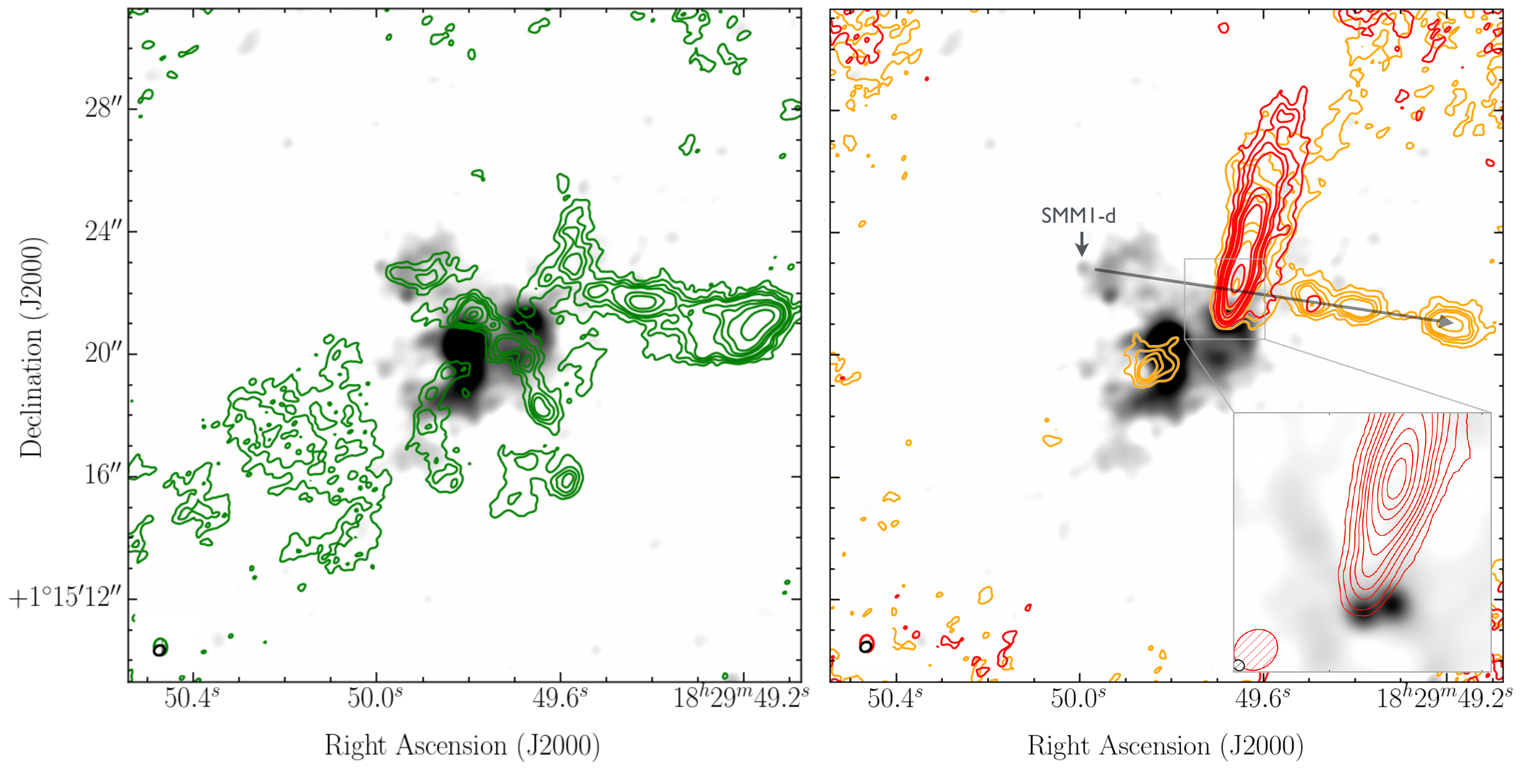}
\caption{\footnotesize   
\textit{Left:} Moment 0 map of \sio (green contours) overlaid on ALMA 1.3\,mm dust continuum emission (grayscale, from ALMA project 2013.1.00726.S). The moment 0 map is constructed by integrating emission from --0.6 to 0.8 \kms with respect to the \vlsr of $\sim$\,8.5\,\kms; contours are 3,\,5,\,7,\,9,\,15,\,20,\,28,\,50 $\times$ the rms noise level of 4.3\,\mjybmkms.  The 1.3\,mm emission peaks at 330\,\mjybm and has an rms noise level of 0.5\,\mjybm. 
\textit{Right:} same as the left panel but for moment 0 maps integrated over different velocity bins: 5.5--25.3\,\kms (orange) and 25.4--39.9\,\kms (red). Contours are the same as on the left for rms noise values of 30 and 26\,\mjybmkms for the orange and red contours, respectively.  The arrow indicates that SMM1-d is the origin of the low-velocity, E--W outflow.  The synthesized beam of the SiO map is $0\farcs55 \times 0\farcs43$ at a position angle of $5\degr$.  The (smaller) synthesized beam of the dust map is $0\farcs37\,\times\,0\farcs31$ at a position angle of --59$\degr$.  
\textit{Right inset:}
Moment 0 map of SiO ($J = 5 \rightarrow 4$) (red contours) overlaid on ALMA 1.3\,mm dust continuum emission (grayscale, from ALMA project 2015.1.00354.S; Pokhrel et al., in prep.). The map is constructed by integrating emission from 25.4--39.9\,\kms with respect to the \vlsr of $\sim$\,8.5\,\kms.  The contours are 3,\,6,\,8,\,11,\,13,\,15,\,17,\,20,\,28,\,35,\,40,\,45 $\times$ the rms noise level of 18\,\mjybmkms. The continuum emission peaks at 14\,\mjybm and has an rms noise level of 140\,\ujybm.  The SiO map was imaged with robust = --1 weighting.  The synthesized beam of the 1.3\,mm continuum map is $0\farcs11 \times 0\farcs10$ at a position angle of 43$\degr$. The synthesized beam of the SiO map is $0\farcs35 \times 0\farcs31$ at a position angle of --5$\degr$.
} 
\vspace{0.2in}
\label{fig:sio} 
\end{figure*}

\begin{figure} [hbt!]
\centering
\includegraphics[scale=0.4, clip, trim=0cm 0cm 0cm 0cm]{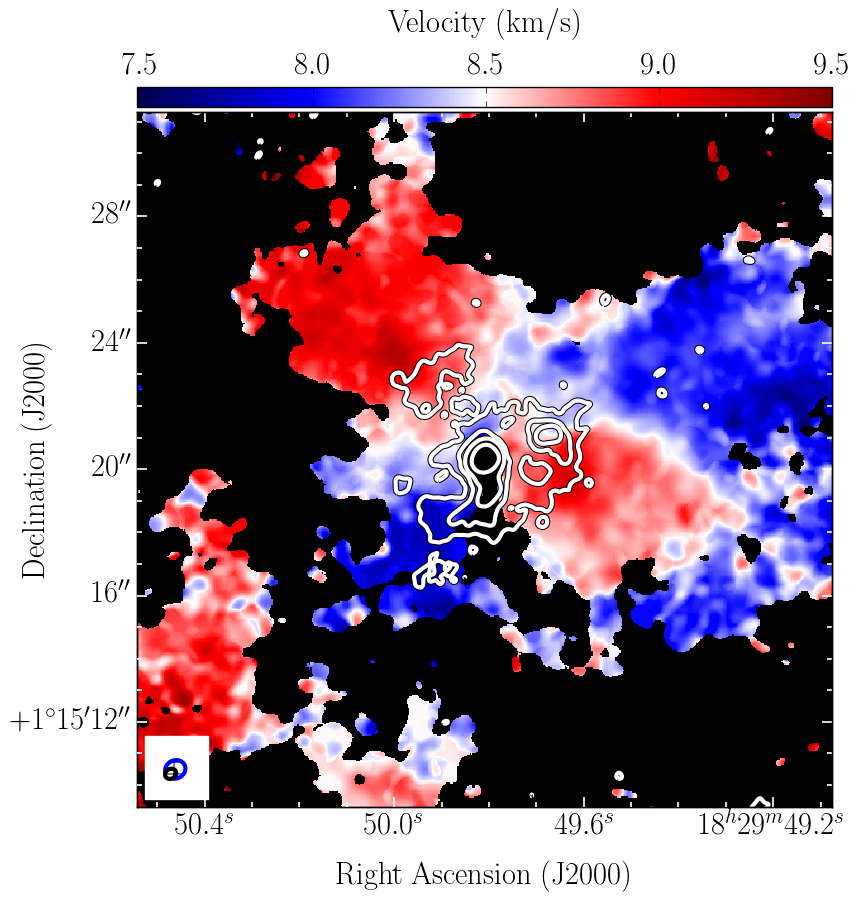}
\caption{\footnotesize   
Moment 1 \dco map (color scale) with overlaid map of ALMA 1.3\,mm dust emission (gray contours). The moment 1 map is constructed from DCO$^+$ spectra integrated from --2 to 2 \kms with respect to the \vlsr of $\sim$\,8.5\,\kms, and was imaged using \textit{uv}-distances $<$\,400\,k$\lambda$ in order to increase the sensitivity to the larger scales. Pixels below 2\,$\times$ the rms noise level of 5.7\,\mjybm are masked. The diverging color scale has been set such that the white color represents the \vlsr.  White contours are 4,\,12,\,26,\,124 $\times$ the rms noise level in the 1.3\,mm dust continuum map of 0.5\,\mjybm.  The synthesized beam of the DCO$^+$ map is $0\farcs67\,\times\,0\farcs59$ at a position angle of --65$\degr$.  The (smaller) synthesized beam of the dust map is $0\farcs37\,\times\,0\farcs31$ at a position angle of --59$\degr$.
} 
\vspace{0.2in}
\label{fig:dco+} 
\end{figure}

In order to put into context the magnetic field morphology with the kinematic properties of the molecular gas, here we present a selected set of molecular emission maps from ALMA: \co (Figure \ref{fig:alma_pol_co}), low- and high-velocity \sio (Figure \ref{fig:sio}), and \dco (Figure \ref{fig:dco+}).  The CO and high-velocity SiO emission trace the molecular outflows/jets emanating from the protostars; the low-velocity SiO emission traces extended material experiencing low-velocity shocks or photodesorption of grains' ice mantles by UV radiation; and the DCO$^+$ traces the dense gas in which the protostars are embedded.

Serpens SMM1 is known to be associated with two high-velocity molecular jets powered by SMM1-a and SMM1-b \cite[][and references therein]{Hull2016a}.  The outflow from SMM1-a has a low-velocity component detected in \co (see Figure \ref{fig:alma_pol_co}); these results are in agreement with the outflow detected by CARMA in \citet{Hull2014}, and with single-dish \cothreetotwo observations out to $\sim$\,1$\arcmin$ scales \citep{Dionatos2010}.  The outflow also coincides with the orientation of the radio jet powered by SMM1 \citep{Curiel1993}.  

SMM1-a and SMM1-b both have extremely high velocity, highly collimated molecular jets.  A high-velocity \co jet emanating from SMM1-a was reported in \citet{Hull2016a}.  In Figure \ref{fig:sio} we report a high-velocity \sio jet emanating from SMM1-b, the companion to the west of SMM1-a.  Furthermore, using 1.3\,mm ALMA dust continuum data with $\sim$\,$0\farcs1$ resolution (Pokhrel et al., in prep.), we show that SMM1-b is a binary with a separation of $\sim$\,$0\farcs3$ ($\sim$\,130\,AU), and that the high-velocity, one-sided SiO jet is driven by the eastern member of the binary.  Highly asymmetric, one-sided outflows have been seen before \citep[e.g.,][]{Pety2006, Loinard2013, Kristensen2013a, Codella2014}; the origin of the asymmetry is unknown, but it may offer important clues about outflow launching mechanisms or the distribution of ambient material near the driving source.

Neither the high-velocity \co jet \citep{Hull2016a} nor the high-velocity \sio jet (Figure \ref{fig:sio}, right panel) exhibits an obvious relationship with the magnetic field in SMM1.  However, the redshifted lobe of the low-velocity \co outflow is clearly shaping the magnetic field morphology (see Figure \ref{fig:alma_pol_co}).  See Section \ref{sec:outflow_shaping} for further discussion.

The low-velocity SiO reveals a new, highly collimated, redshifted outflow oriented roughly E--W direction (Figure \ref{fig:sio}). Its axis points clearly toward the faintest source we detect, SMM1-d. Thus, SMM1-d is likely to be a previously undetected low mass protostar. SMM1-c is the only compact source in the region that does not show clear outflow activity.

We analyze \dco emission to better understand the kinematics of the dense material in the envelope surrounding SMM1-a and SMM1-b.  DCO$^+$ traces the dense,  $\sim$\,20--30\,K molecular gas\footnote{In order for DCO$^+$ to be present, the temperature must be cold enough for deuterium chemistry to be active, but not so cold that CO is depleted onto dust grains.  See \citet{Jorgensen2011}.}  around the protostars at scales ranging from a few $\times$ 100\,AU up to a few $\times$ 1000\,AU.  The line emission shows smooth (and seemingly quadrupolar) velocity gradients of $\sim$1.0\,\kms within a scale of $\sim$\,1000\,AU.  However, the gradients, while relatively ordered, have little correlation with the magnetic field or outflow orientations.  

Finally, we analyze extended \sio emission near the systemic velocity of SMM1.  Narrow-line-width SiO emission at systemic velocities has been detected toward very dense regions around protostars \citep[e.g.,][]{Girart2016}. This type of emission may be due to the presence of low-velocity shocks \citep{JimenezSerra2010, Nguyen2013}; however, extended SiO emission near the systemic velocity can also be caused by photodesorption of SiO from dust grains' icy mantles by UV radiation (see Appendix B of \citealt{Coutens2013}, and references therein).  The low-velocity SiO emission toward SMM1 is patchy, and is spread out across the field of view. While the strongest emission is associated with the E--W SiO outflow mentioned above, the SiO that is spatially coincident with the dust emission has a distinctive $\sim$\,3000\,AU arc-like ridge that passes through the lower density region between SMM1-a and SMM1-b.  This emission is located in a region with significant depolarization in some places, and a chaotic magnetic field in the regions where polarization is detected.  Assuming the emission comes from low-velocity shocks, this suggests that the magnetic field may have been perturbed by a bow-shock front that is crossing the dense core. The large scale of this front suggests an external origin, e.g., from large-scale turbulence; this is consistent with the complex dynamics of Serpens Main \citep{Lee2014}, which may have formed in a cloud-cloud collision \citep{DuarteCabral2011}. 

For channel maps and a brief discussion of other dense molecular tracers detected toward SMM1 by the SMA, see Appendix \ref{appendix:lines}.

\section{Discussion}
\label{sec:dis}

\subsection{Magnetic fields at different spatial scales}

Optical polarization and (sub)millimeter observations have revealed that magnetic fields at large ($\gtrsim$\,1\,pc) scales tend to be relatively uniform and correlated with the molecular cloud morphology \citep{Pereyra2004, HBLi2006, Alves2008, Goldsmith2008, Franco2010, Palmeirim2013, Fissel2016}.  The magnetic fields seem to have a bimodal behavior, where the field is either parallel or perpendicular to the major axis of the cloud \citep{HBLi2009, HBLi2013, Soler2013, PlanckXXXII, PlanckXXXV}.  This orderliness and bimodality of the magnetic fields is also observed at the $\sim$\,0.1--0.01\,pc protostellar core scale \citep{Koch2014, Zhang2014}.  In addition, recent studies in the NGC 6334 cloud show that the mean magnetic field orientation does not change significantly between $\sim$\,100\,pc and $\sim$\,0.01\,pc scales \citep{HBLi2015}. These observational results agree with simulations of magnetically regulated evolution of molecular clouds \citep{Kudoh2007, Nakamura2008, Tomisaka2014}. 

In Serpens SMM1 at $\gtrsim$\,0.1\,pc scales, near-infrared and submillimeter polarization maps show that the magnetic field is perpendicular to the filamentary structure seen in the dust emission \citep{Davis1999, Davis2000, Matthews2009, Sugitani2010}, as observed in many other regions, such as some of those listed above.  However, Figure \ref{fig:pol} shows that within the core, the magnetic field as traced by CARMA and the SMA appears significantly perturbed, especially compared with the larger-scale component.  The dramatic change in the magnetic field configuration between 0.1--0.01\,pc does not fit with the aforementioned properties of magnetic fields in molecular clouds and cores. 

This change in magnetic field orientation from 0.1--0.01\,pc scales is not unique, and is seen in both high mass sources (e.g., DR21(OH); see \citealt{Girart2013}) and many low-mass sources \citep{Hull2014}.  Specifically, our SMM1 results can be compared with the ALMA polarization observations of Ser-emb 8, another Class 0 protostellar source in the Serpens Main cloud \citep{Hull2017a}.  After analyzing the observations in concert with high-resolution MHD simulations, \citeauthor{Hull2017a} argued that the inconsistency of the magnetic field orientation across several orders of magnitude in spatial scale in Ser-emb 8 may be because the source formed in a highly turbulent, weakly magnetized environment.  This may be true for SMM1 as well; however, unlike Ser-emb 8, SMM1 shows clear evidence that the outflow has shaped the field at the small scales observable by ALMA.  Below we discuss this and other effects that can help us understand the changes in the magnetic field orientation across multiple spatial scales in SMM1.

\subsection{Shaping of the magnetic field by the wide-angle, low-velocity outflow from SMMI-a}
\label{sec:outflow_shaping}

It is clear from Figure \ref{fig:alma_pol_co} that the magnetic field to the SE of SMM1-a is being shaped by the wide-angle, low velocity \co outflow.  In fact, the magnetic field also appears to trace the base of the blueshifted outflow lobe, although there are many fewer independent detections of polarization on that (NW) side of the source (see Section \ref{sec:MolEm}).  However, while the low-velocity CO outflow corresponds well with the magnetic field morphology toward SMM1, the high-velocity jet components do not.  \citet{Hull2016a} studied the EHV CO jet emanating to the SE of SMM1-a, which seems to bisect the $\sim$\,90$\degree$ opening created by the low-velocity outflow, but does not obviously shape the magnetic field lying along either cavity wall.  Furthermore, in Figure \ref{fig:sio} we show redshifted EHV SiO emission from SMM1-b, which does not obviously shape the magnetic field toward that source.

Why the magnetic field in SMM1 is shaped by the low-velocity outflow but not the high-velocity jet is an open question.  In the case of SMM1-a, the wide-angle cavity has probably been excavated by the low-velocity outflow, leaving little material with which the narrow, high-velocity CO jet can interact.  At the same time, the pressure from the outflow increases the column density (and possibly compresses the magnetic field) along the edges of the cavity; this allows us to detect the effects of the outflow on the magnetic field pattern because the column density (and thus the brightness of the optically thin polarized and unpolarized dust emission) is highest at the cavity edge.  However, in the case of SMM1-b, which has no wide-angle outflow, the narrow SiO jet (and the corresponding EHV CO jet from \citealt{Hull2016a}) still does not have an obvious effect on the magnetic field, suggesting that perhaps the solid angle of material being affected by the jet is simply too small to be seen in the ALMA polarization maps.  

Note that we may see more prominent sculpting of the magnetic field toward SMM1-a because it may be more evolved than SMM1-b, and thus has a wider outflow cavity.  Some studies have found a correlation between outflow opening angle and protostellar age, where older sources have wider outflows \citep{ArceSargent2006}.  However, more recent infrared scattered-light studies have come to a variety of conclusions, suggesting that the relationship between outflow opening angle and age is not yet certain \citep{Seale2008, Velusamy2014, Booker2017, Hsieh2017}.

\subsection{Energetics estimates}
\label{sec:energetics}

While it seems reasonable to assume that the outflow has shaped the magnetic field in SMM1-a, it is nonetheless prudent to compare the importance of the three main effects that can shape the magnetic field at the small spatial scales we are probing with the ALMA observations: namely, the outflow, the magnetic field, and gravity.  One motivation for making these comparisons is that the magnetic field within the inner $\sim$\,500\,AU of the source (as revealed by the ALMA data in Figure \ref{fig:pol}(d)) does seem to resemble a small hourglass with its axis along the outflow axis (see the discussion of hourglass-shaped fields in Section \ref{sec:intro}).  A comparison of the magnetic vs. outflow energy can shed light on whether this hourglass-shaped magnetic field immediately surrounding SMM1-a is part of a strongly magnetized preexisting envelope that has shaped the outflow; or whether, as we assume above, that the outflow has shaped the magnetic field and the hourglass shape is simply tracing the base of the outflow cavity.

\subsubsection{Gravitational potential energy}

To estimate the gravitational potential energy we must first estimate the mass of the dust measured by ALMA toward SMM1.  The ALMA map pictured in Figure \ref{fig:pol}(d) has a total 343\,GHz Stokes $I$ flux density $S_{\nu} \sim 4.6$\,Jy within a circle of radius 4$\arcsec$, or $\sim$\,1700\,AU, centered on the peak of SMM1-a.  However, the dust nearest to SMM1-a and SMM1-b is likely to be significantly warmer.  Thus, we separate the map into three regions: (1) a region immediately surrounding SMM1-a with a flux of $\sim$\,2.2\,Jy, (2) a region immediately surrounding SMM1-b with a flux of $\sim$\,0.3\,Jy, and (3) the rest of the region, with a flux of 2.1\,Jy.  We assume dust temperatures $T_d \sim 50$\,K for the dust near SMM1-a and b, and $T_d \sim 20$\,K for the remainder of the dust.\footnote{The $\sim$\,20\,K value for the dust not in the immediate vicinity of the protostars is based on an estimate provided by Katherine Lee (2015, private communication).  That value was from a dust temperature map of Serpens that was derived from spectral energy distribution (SED) fits to \textit{Herschel} maps; the same method was used by \citet{Storm2016} to estimate temperatures in the L1451 star-forming region, and is described in Section 7.1 of that publication.  In all cases, the \textit{Herschel} zero-point fluxes had been corrected using \textit{Planck} maps, as described in \citet{MeisnerFinkbeiner2015}.}

We convert the flux $S_{\nu}$ contained within the area under consideration into a corresponding gas mass estimate using the following relation:

\begin{equation}
M_{\mathrm{gas}}=\frac{S_{\nu}d^{2}}{\kappa_{\nu}B_{\nu}\left(T_{\mathrm{d}}\right)}\,\,.
\label{eq:M_gas}
\end{equation}

\noindent 
$B_{\nu}\left(T_{\mathrm{d}}\right)$ is the Planck function at the frequency of the observations.  Using a distance $d=436$\,pc and an opacity $\kappa_{\nu}=2\,\mathrm{cm}^2/\mathrm{g}$ \citep{Ossenkopf1994}, and assuming a gas-to-dust ratio of 100, we obtain a combined gas mass in all three regions of $M_{\mathrm{gas}}\approx3.8\,M_\sun$.\footnote{Note that we assume that all of the dust is optically thin; this may not be true very close to SMM1-a, which would result in an underestimate of the gas mass.} Using a radius of 1700\,AU, this quantity can be converted into a mean gas volume density $\rho \sim 1\times10^{-16}\,\mathrm{g}/\mathrm{cm}^{-3}$ and mean gas number density $n \sim \,2.9 \times10^7\, \mathrm{cm}^{-3}$ (assuming a mean molecular mass of $2.3$).  

% Volume density calculation (this DOES NOT include the star!)
% 3.8 solar masses / (4/3 * pi * (1700 AU)^3) in g * cm^(-3)

To calculate the mass of SMM1-a, the most massive protostar in the system, we use mass-luminosity relations for pre-main-sequence stars \citep{Yorke2002} and find that a protostar with the luminosity of SMM1-a ($L \sim 100\,L_{\odot}$) has a mass of $\sim$\,3\,$M_{\odot}$.

Using a total mass of 6.8\,$M_\odot$ and a radius of 1700\,AU, we calculate a gravitational potential energy of $E_\textrm{grav} \sim$\,$4.8 \times 10^{44}$ erg.

% GRAV PE CALCULATION
% G * (6.8 solar mass)^2 / (1700 AU) in ergs

% Planck function prefactor at 343 GHz:
% 2*h*(343 GHz)^3/c^2 in g/s^2 = 5.95e-13 g/s^2

% Planck function exponential for *** 50K *** at 343 GHz:
% e^(2.27274202e-22 / 6.9032426e-22) - 1 = 0.3899

% TOTAL PLANCK = 1.52e-12 g/s^2

% Planck function exponential for *** 20K *** at 343 GHz:
% e^(2.27274202e-22 / 2.76129704e-22) - 1 = 1.277

% TOTAL PLANCK = 4.659e-13 g/s^2

%%% THREE REGIONS
% (1) SMM1-a (50 K, 2.2 Jy)
%
%2.2 Jansky * (436 parsec)^2 / (0.02 cm^2/g) / 1.52e-12 g/s^2
%(2.2 * 1e-23) * (436 * 3.086e18)^2 / 0.02 / 1.52e-12 / 1.33e33
%= 0.66 Msun
%
%% (2) SMM1-b (50 K, 0.3 Jy)
%
%(0.3 * 1e-23) * (436 * 3.086e18)^2 / 0.02 / 1.52e-12 / 1.33e33
%= 0.09 Msun
%
%% (3) rest of region (20 K, 2.1 Jy)
%
%(2.1 * 1e-23) * (436 * 3.086e18)^2 / 0.02 / 4.659e-13 / 1.33e33
%= 3.08 Msun

\subsubsection{Magnetic field energy}

Our calculations for the magnetic field strength follow the procedure outlined in \citet{Houde2016}. Specifically, we calculate the dispersion in polarization angles from the ALMA polarization map using the function $1-\left\langle \cos\left[\Delta\Phi\left(\ell\right)\right]\right\rangle$, where the quantity $\ell$ is the distance between a pair of polarization orientations. The dispersion due to the turbulent component of the magnetic field is isolated for the analysis by removing the large-scale component, which comprises a constant term and a second-order term (in $\ell$); this yields a turbulence correlation length of $\delta \simeq 0.3\arcsec$. The effective thickness of the cloud is assumed to be similar to its extent on the sky and is estimated from the width of the autocorrelation function of the polarized flux ($\Delta^{\prime} \simeq 0.44\arcsec$). The combination of $\delta$ and $\Delta^{\prime}$ with the width of the ALMA synthesized beam implies that, on average, approximately one turbulent cell is contained in the column of gas probed by the telescope beam.  The resulting turbulent-to-total magnetic energy ratio $\left\langle B_{\mathrm{t}}^{2}\right\rangle / \left\langle B^{2}\right\rangle = 0.25$ \citep{Hildebrand2009, Houde2009, Houde2016}. This quantity is then used with both the mean volume density $\rho$ calculated above as well as the one-dimensional turbulent velocity dispersion $\sigma\left(v\right)\sim 0.8 \:\mathrm{km}\:\mathrm{s}^{-1}$ (from our unpublished $^{13}\mathrm{CS}\left(v=0, 5\rightarrow4\right)$ ALMA data toward this source) to calculate a magnetic field strength of $\sim 5.7$\,mG (plane-of-the-sky component) with the so-called Davis-Chandrasekhar-Fermi equation \citep{Davis1951, Chandrasekhar1953}:

\begin{equation}
B_{0}\simeq\sqrt{4\pi\rho}\,\sigma\left(v\right)\left[\frac{\left\langle B_{\mathrm{t}}^{2}\right\rangle }{\left\langle B^{2}\right\rangle }\right]^{-1/2}.\label{eq:DCF}
\end{equation}

% B-field calculation
% sqrt(4*pi*1e-16) * (0.8 * 1e5) * 1/sqrt(0.25)

\noindent
Given the energy density of the magnetic field $B^2 / 8 \pi$ and a radius of 1700\,AU, we calculate the magnetic energy in the material surrounding SMM1 to be $E_B \sim$\,$9 \times 10^{43}$\,erg.

% B-FIELD ENERGY CALCULATION (whole sphere, not just 1/4)
% (4/3) * pi * (1700 * 1.496e13)^3 * (5.7e-3)^2/(8*pi)

\subsubsection{Outflow energy}

Following the methods outlined in \citet{Zhang2001, Zhang2005}, we calculate the energy in the redshifted lobe of the \co outflow launched by SMM1 using both the ALMA data presented here as well as the CARMA data presented in Figure 27 of \citet{Hull2014}.  We assume a distance of 436\,pc, a temperature of 20\,K, and optically thin emission.  We do not correct for the inclination of the outflow.  Analysis of the CARMA data yields a total redshifted outflow mass $M_\textrm{out} = 0.03\,M_\odot$, momentum $P_\textrm{out} = 0.29$\,$M_\odot$\,\kms, and energy $E_\textrm{out} = 1.53\,M_\odot$\,(\kms)$^2$.  The ALMA values are $M_\textrm{out} = 0.006\,M_\odot$, $P_\textrm{out} = 0.021$\,$M_\odot$\,\kms, and $E_\textrm{out} = 0.061\,M_\odot$\,(\kms)$^2$.  The values calculated from the ALMA data are significantly lower because ALMA is unable to recover a substantial fraction of the large-scale emission from the outflow.  It is worth noting that the values calculated from the CARMA data are comparable to the results obtained by \citet{Davis1999}, who used JCMT (single-dish) data to measure the energetics for the aggregate sample of outflows in the Serpens Main region.  Thus, for the purposes of this energetics analysis, we adopt the CARMA value of $E_\textrm{out} = 1.53\,M_\odot$\,(\kms)$^2$, or $3 \times 10^{43}$\,erg.  

% OUTFLOW ENERGY CALCULATION
% 1.53 solar mass * (1 km/s)^2 in ergs 

\subsubsection{Energy comparison}

The redshifted lobe of the outflow pictured in Figure \ref{fig:alma_pol_co} has an opening angle of approximately 90$\degree$ in the region of interest, and thus occupies $\sim$\,$\frac{1}{7}$ of the volume of the sphere surrounding SMM1-a that we use in the magnetic and gravitational energy estimates above. Scaling the magnetic and gravitational energies down by a factor of 7 to compare with the outflow energy $E_\textrm{out} \sim 3 \times 10^{43}$\,erg, we find $E_B \sim$\,$1.3 \times 10^{43}$\,erg and $E_\textrm{grav} \sim$\,$6.9 \times 10^{43}$ erg.  

% SCALED DOWN NUMBERS
%
% GRAV: 4.8e44 / 7 = 6.9e43
%
% MAG: 9e43 / 7 = 1.3e43

In summary, the gravitational, magnetic, and outflow energies are all comparable.  There is substantial uncertainty in several of the parameters that go into the above estimates: the outflow energy derived from the CARMA data is a lower limit on the true value because of the interferometer's inability to recover emission at all spatial scales; the dust temperature and optical depth at high resolution are not well constrained; and \cs may or may not be the best species to use to estimate the turbulent line width for the DCF magnetic field estimate.  Consequently, while the numbers do not allow us to make a strong claim that either the outflow or the magnetic field is dominant in SMM1, we nonetheless find our assumption---that the outflow may have shaped the magnetic field---to be reasonable.

\subsection{Biased polarization images due to beam smearing}
\label{sec:pol_bias}

Figures \ref{fig:pol} and \ref{fig:alma_pol_co} show that the magnetic field follows the edge of the outflow cavity traced by the low-velocity, redshifted CO emission emanating to the SE of SMM1-a.  However, the intensity of the polarized emission is very different on the two sides of the cavity: the E--W component is several times weaker than the N--S component.  With ALMA we are able to resolve the two components fully; however, previous observations by CARMA and the SMA (see Figure \ref{fig:pol}) had 5--10 times lower resolution, which led these two components to be blended together, with the N--S component clearly dominating.

\begin{figure*} [hbt!]
\centering
\includegraphics[width=\linewidth, clip, trim=0cm 0cm 0cm 0cm]{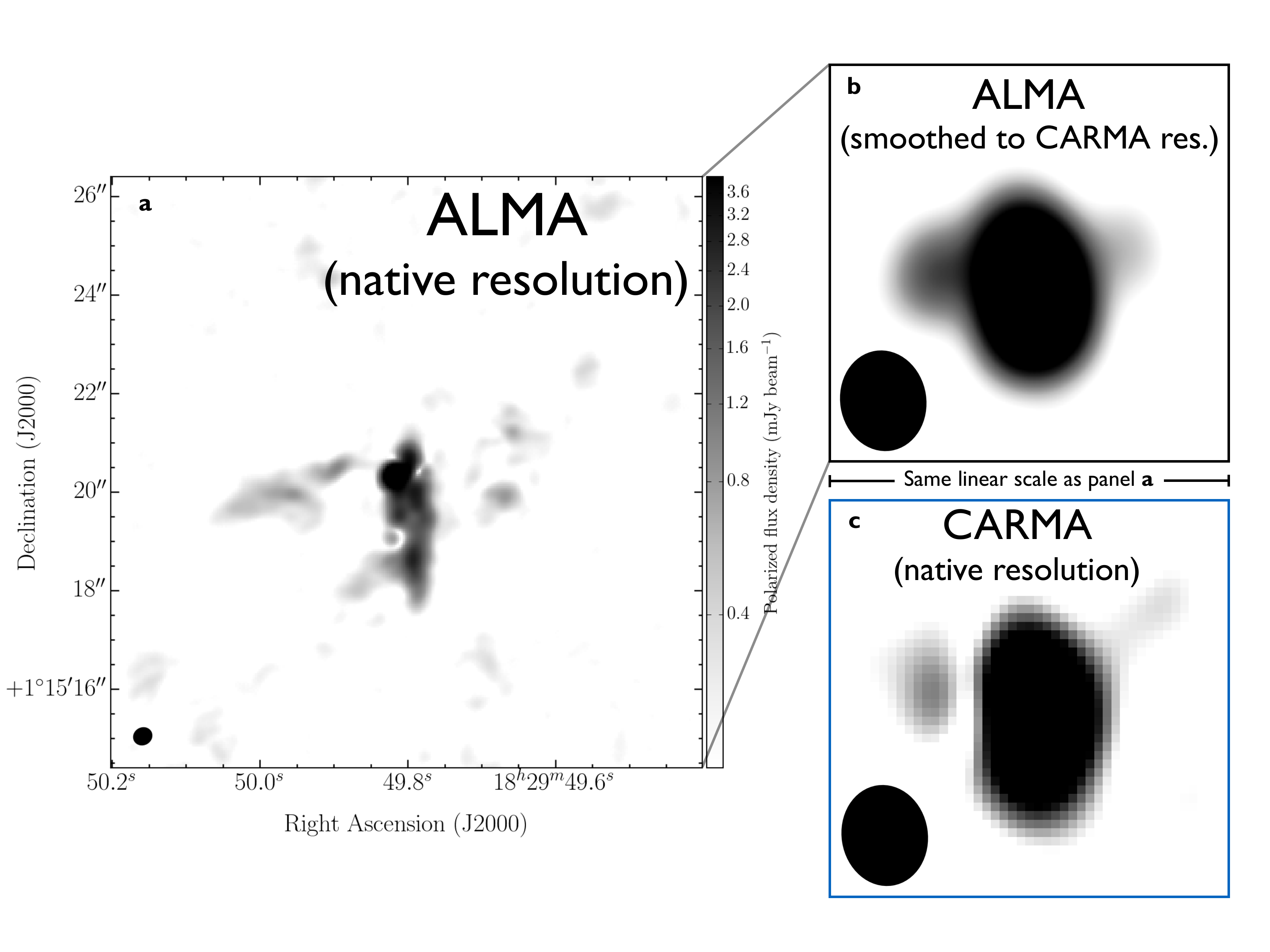}
\caption{\footnotesize   
Maps of the polarized intensity toward SMM1.  Panel (a) shows the ALMA 870\,\micron{} image of polarized dust emission at the native resolution of $0\farcs3$.   While the peak polarized intensity of the ALMA image is 11.8\,\mjybm, the color scales in all panels have been saturated to enhance the low-level structure (hence the reason why the color bar maximum is $\sim$\,3.6\mjybm).  Panel (b) shows the smoothed ALMA data, where the image was produced by tapering the \textit{uv} data and smoothing the image to match the $\sim$\,2.5$\arcsec$ native resolution of the CARMA image, shown in panel (c).  Note that the ALMA map in panel (b) looks much smoother than the CARMA map simply because the pixel size is smaller.
} 
\vspace{0.2in}
\label{fig:smm1_poli} 
\end{figure*}

In Figure \ref{fig:smm1_poli} we show polarized intensity maps from both CARMA and ALMA.  The CARMA data are at their original resolution (Figure \ref{fig:smm1_poli}(b)), whereas the ALMA data are tapered and smoothed to produce a map with the same resolution (Figure \ref{fig:smm1_poli}(c)).  The similarity is striking: when the ALMA data are smoothed to CARMA resolution, the E--W component is dwarfed by the much brighter N--S component.  It is thus clear that we must proceed with caution when revisiting low-resolution polarization maps, as plane-of-sky beam smearing biases the maps in favor of the material with the brightest polarized emission.

\subsection{Gravitational infall or dust scattering}

In the region immediately surrounding SMM1 (within a few $\times$ 100\,AU; see the inner few resolution elements of Figure \ref{fig:pol}(d)), the magnetic field orientation looks somewhat radial, which could indicate that the field lines are being dragged in by gravitational collapse, similar to the radial magnetic field configuration that was seen in SMA observations of the high-mass star-forming core W51~e2 \citep{Tang2009b}.
A radial magnetic field pattern is derived from an azimuthal polarization pattern, assuming that the polarization arises from magnetically aligned dust grains (i.e., the magnetic field orientations are perpendicular to the polarization orientations, as was assumed in Figures \ref{fig:pol} and \ref{fig:alma_pol_co} and described in Section \ref{sec:intro}).  However, an azimuthal polarization pattern can also arise from self-scattering of dust emission from a face-on (or slightly inclined) protoplanetary disk: recent theoretical work has shown that, depending on the combination of dust density, dust-grain growth, optical depth, disk inclination, and resolution of observations, polarization from scattering in disks could contribute to the polarized emission at millimeter wavelengths, perhaps even eclipsing the signal from magnetically aligned dust grains \citep{Kataoka2015, Kataoka2016, Pohl2016, Yang2016a, Yang2016b, Yang2017}.
There is now potential evidence for this dust scattering effect from ALMA observations \citep{Kataoka2016b}; other high-resolution polarization observations by CARMA and the Karl G. Jansky Very Large Array (VLA) \citep{Stephens2014,Cox2015,FernandezLopez2016} may also be consistent with self-scattered dust emission.  However, while intriguing, our current data do not allow us to resolve the disk sufficiently well to differentiate between the two scenarios described above.  We will further investigate this question of magnetic fields vs. scattering with higher-resolution ALMA polarization observations of SMM1 (Hull et al., in prep.).  

Note that in order for scattering of dust emission to be efficient at (sub)millimeter wavelengths, the grains must be of order a few $\times$ 100\,$\micron$ \citep{Kataoka2015}.  While scattering may be important toward the very center of SMM1, it is highly unlikely that scattering is the dominant effect at scales $\gtrsim$\,100\,AU where grains are expected to be a few microns in size.  Therefore, nearly all of the polarized emission in all panels of Figure \ref{fig:pol} is likely to be produced by magnetically aligned dust grains, especially if the emitting grains reside in a rapidly infalling envelope (as opposed to a rotationally supported disk), where grains are unlikely to grow to hundreds of microns because of the short dynamical timescale and relatively low density of the material.

%%%%%%%%%%%%%%%%%%%%%%%%%%%%%%%%%%
% TABLE: ALMA polarization data

\begin{table}[hbt!]
\normalsize
\begin{center}
\caption{\normalsize \vspace{0.1in} ALMA polarization data}
\begin{tabular}{cccccc}
\hline
\hline
$\alpha_{\textrm{J2000}}$ & $\delta_{\textrm{J2000}}$ & $\chi$ & $\delta\chi$ & $P$ & $I$\\
(\degree) & (\degree) & (\degree) & (\degree) & \mjybmvert & \mjybmvert \\
\hline
277.45868 &   1.25424 &    86.5 &     6.9 &   0.250 &     --- \\ 
277.45862 &   1.25424 &    95.7 &     6.8 &   0.254 &     --- \\ 
277.45857 &   1.25424 &    98.3 &     9.4 &   0.182 &     --- \\ 
277.45868 &   1.25429 &    97.9 &     9.3 &   0.185 &     --- \\ 
277.45862 &   1.25429 &   104.6 &     7.0 &   0.246 &     --- \\ 
277.45857 &   1.25429 &   115.5 &     7.5 &   0.230 &     --- \\ 
277.45673 &   1.25429 &     0.7 &     8.8 &   0.196 &     --- \\ 
277.45612 &   1.25429 &    27.4 &     9.5 &   0.181 &     --- \\ 
277.45896 &   1.25435 &   123.1 &     9.0 &   0.192 &     --- \\ 
277.45873 &   1.25435 &   128.0 &     8.7 &   0.197 &     --- \\ 
277.45718 &   1.25435 &    84.3 &     8.3 &   0.207 &     --- \\ 
277.45712 &   1.25435 &    76.9 &     5.6 &   0.304 &     --- \\ 
277.45707 &   1.25435 &    65.7 &     8.2 &   0.209 &     --- \\ 
277.45634 &   1.25435 &    53.8 &     8.8 &   0.195 &     --- \\ 
277.45896 &   1.25441 &   133.4 &     6.6 &   0.261 &     --- \\ 
277.45840 &   1.25441 &   134.4 &     9.4 &   0.182 &     --- \\ 
277.45712 &   1.25441 &    64.9 &     9.4 &   0.182 &     --- \\ 
277.45696 &   1.25441 &    47.4 &     7.5 &   0.230 &     --- \\ 
277.45896 &   1.25446 &   137.6 &     4.9 &   0.351 &     --- \\ 
277.45890 &   1.25446 &   142.4 &     7.9 &   0.217 &     --- \\ 
277.45846 &   1.25446 &   136.7 &     8.4 &   0.204 &     --- \\ 
277.45840 &   1.25446 &   138.7 &     9.2 &   0.187 &     --- \\ 
277.45834 &   1.25446 &   136.5 &     8.0 &   0.215 &    1.664 \\ 
277.45701 &   1.25446 &    15.2 &     8.9 &   0.193 &     --- \\ 
277.45696 &   1.25446 &    30.5 &     7.4 &   0.231 &     --- \\ 
277.45896 &   1.25452 &   142.9 &     5.9 &   0.291 &     --- \\ 
277.45890 &   1.25452 &   143.9 &     6.0 &   0.288 &     --- \\ 
277.45834 &   1.25452 &   140.5 &     6.9 &   0.247 &    2.468 \\ 
277.45829 &   1.25452 &   138.4 &     9.0 &   0.192 &    2.094 \\ 
277.45707 &   1.25452 &   158.2 &     7.0 &   0.247 &     --- \\ 
277.45701 &   1.25452 &   170.5 &     6.4 &   0.270 &     --- \\ 
277.45696 &   1.25452 &     6.3 &     8.7 &   0.199 &     --- \\ 
277.45896 &   1.25457 &   150.3 &     5.9 &   0.290 &     --- \\ 
277.45890 &   1.25457 &   157.5 &     6.3 &   0.271 &     --- \\ 
277.45884 &   1.25457 &   170.9 &     7.8 &   0.219 &     --- \\ 
277.45829 &   1.25457 &   134.5 &     8.0 &   0.215 &    2.933 \\ 
277.45723 &   1.25457 &   101.3 &     8.4 &   0.205 &     --- \\ 
277.45712 &   1.25457 &   131.3 &     7.3 &   0.235 &     --- \\ 
277.45707 &   1.25457 &   134.4 &     5.0 &   0.345 &     --- \\ 
277.45701 &   1.25457 &   143.5 &     6.1 &   0.284 &     --- \\ 
277.45690 &   1.25457 &   175.3 &     9.4 &   0.182 &     --- \\ 
277.45646 &   1.25457 &   135.1 &     8.5 &   0.202 &     --- \\ 
277.45896 &   1.25463 &   149.3 &     6.8 &   0.252 &     --- \\ 
... &   ... &   ... &     ... &   ... &    ... \\ \\
\hline
\vspace{-1.5em}
\end{tabular}
\label{table:data}
\end{center}
\footnotesize
\textbf{Note.}  $\chi$ is the orientation of the magnetic field, measured counterclockwise from north.  $\delta\chi$ is the uncertainty in the magnetic field orientation.  $P$ is the polarized intensity.  $I$ is the total intensity, reported where $I > 3\,\sigma_I$.  Due to differences in dynamic range between the images of Stokes $I$ and polarized intensity, there are cases where $P$ is detectable but $I$ is not.  
\textit{The full, machine-readable table is available in the online version of this publication.}
\end{table}
%%%%%%%%%%%%%%%%%%%%%%%%%%%%%%%%%%

\section{Conclusions}
\label{sec:con}

We have analyzed the magnetic field morphology toward the Class 0 protostar Serpens SMM1 using new ALMA and SMA polarization data as well as archival CARMA and JCMT polarization data; the combination of these multiple datasets has allowed us to probe spatial scales from $\sim$\,80,000 down to $\sim$\,140\,AU.  We examine the magnetic field morphology in concert with molecular line observations from ALMA and come to the following conclusions:

\begin{enumerate}

\item Dramatic changes in the magnetic field morphology occur between the ``core'' scale of a few $\times$ 0.1\,pc probed by the JCMT and the much smaller ``envelope'' scales probed by the CARMA, SMA, and ALMA interferometers.  These changes are inconsistent with models of strongly magnetized star formation, which predict that the magnetic field orientation should be preserved across many orders of magnitude in spatial scale.

\item Other sources such as Ser-emb~8 \citep{Hull2017a} have shown this multi-scale inconsistency in magnetic field morphology.  However, unlike Ser-emb~8, SMM1 shows a magnetic field morphology that has clearly been affected by its bipolar outflow: the redshifted lobe of the low-velocity \co outflow has excavated a wide-angle cavity, compressing the magnetic field along the cavity edges.  

\item Conversely, the highly collimated, extremely high-velocity CO and SiO jets emanating from SMM1-a and its nearby companion SMM1-b are not obviously shaping the magnetic field.  This suggests that narrow jets do not perturb a large enough fraction of the envelope to have a detectable effect on the magnetic field morphology.  Perhaps SMM1-a is more evolved than sources like SMM1-b or Ser-emb~8, and has entered an evolutionary phase where the magnetic field morphology is shaped by the wider, low-velocity outflow.
\label{conclusions:3}

\item Outside of the region where the magnetic field is shaped by the low-velocity \co outflow emanating from SMM1-a, there appears to be significant depolarization in some places, and a chaotic magnetic field in the regions where polarization is detected.  This may be due to the presence of a large-scale bow shock crossing the envelope and disturbing the magnetic field morphology.

\item Using $\sim$\,$0\farcs1$ resolution ALMA continuum observations, we report that the source SMM1-b is a protobinary with $\sim$\,130\,AU separation.  The eastern component of the binary is powering the extremely high-velocity, one-sided SiO jet mentioned in point \ref{conclusions:3}.

\end{enumerate}

These observations show that with the sensitivity and resolution of ALMA, we can now begin to understand the role that outflow feedback plays in shaping the magnetic field in very young, star-forming sources like SMM1.  Future high-resolution, high-sensitivity ALMA surveys will be necessary to better understand the impact of outflows on the magnetic fields in star-forming cores---in particular, how often protostellar feedback obviously shapes the magnetic field in the natal core, and whether there are correlations between outflow-shaped magnetic fields and source environment, mass, or evolutionary stage.

\acknowledgments
The authors thank the anonymous referee, whose comments improved the manuscript.
C.L.H.H. acknowledges the outstanding calibration and imaging work performed at the North American ALMA Science Center by Crystal Brogan, Jennifer Donovan Meyer, and Mark Lacy.
He also acknowledges Katherine Lee, Shaye Storm, and Aaron Meisner for the helpful discussion regarding dust temperature estimates in Serpens.
The authors thank all members of the SMA staff who made the SMA observations possible.  
J.M.G. is supported by the Spanish MINECO AYA2014-30228-C03-02 and the MECD PRX15/00435 grants.
Q.Z. and J.M.G. acknowledge the support of the SI CGPS award, ``Magnetic Fields and Massive Star Formation,'' and the 
SI SSA grant, ``Are Magnetic Fields Dynamically Important in Massive Star Formation?''
Q.Z. acknowledges the support of SI Scholarly Studies Awards.
Astrochemistry in Leiden is supported by the Netherlands Research School for Astronomy
(NOVA), by a Royal Netherlands Academy of Arts and Sciences (KNAW) professor prize, and by the European Union A-ERC grant 291141 CHEMPLAN.
Z.-Y.L. is supported in part by the NASA NNX14AB38G and NSF AST1313083 grants.
Support for CARMA construction was derived from the states of California, Illinois, and Maryland, the James S. McDonnell Foundation, the Gordon and Betty Moore Foundation, the Kenneth T. and Eileen L. Norris Foundation, the University of Chicago, the Associates of the California Institute of Technology, and the National Science Foundation.
The Submillimeter Array is a joint project between the Smithsonian Astrophysical Observatory and the Academia Sinica Institute of Astronomy and Astrophysics, and is funded by the Smithsonian Institution and the Academia Sinica.
The National Radio Astronomy Observatory is a facility of the National Science Foundation operated under cooperative agreement by Associated Universities, Inc. 
This paper makes use of the following ALMA data: ADS/JAO.ALMA\#2013.1.00726.S and ADS/JAO.ALMA\#2015.1.00354.S.
ALMA is a partnership of ESO (representing
   its member states), NSF (USA) and NINS (Japan), together with NRC
   (Canada), NSC and ASIAA (Taiwan), and KASI (Republic of Korea),
   in cooperation with the Republic of Chile. The Joint ALMA
   Observatory is operated by ESO, AUI/NRAO and NAOJ.
This research made use of APLpy, an open-source plotting package for Python hosted at \url{http://aplpy.github.com}.
The figure in the Appendix was created using the GREG package from the GILDAS data reduction package, available at \url{http://www.iram.fr/IRAMFR/GILDAS}.

\textit{Facilities:} JCMT, CARMA, SMA, ALMA.

%\bibliography{ms}
%\bibliographystyle{apj}

\appendix

\section{SMA observations of dense molecular tracers toward SMM1}
\label{appendix:lines}

\begin{figure*}[hbt!]
\includegraphics[width=\linewidth, clip, trim=0cm 0cm 0cm 0cm]{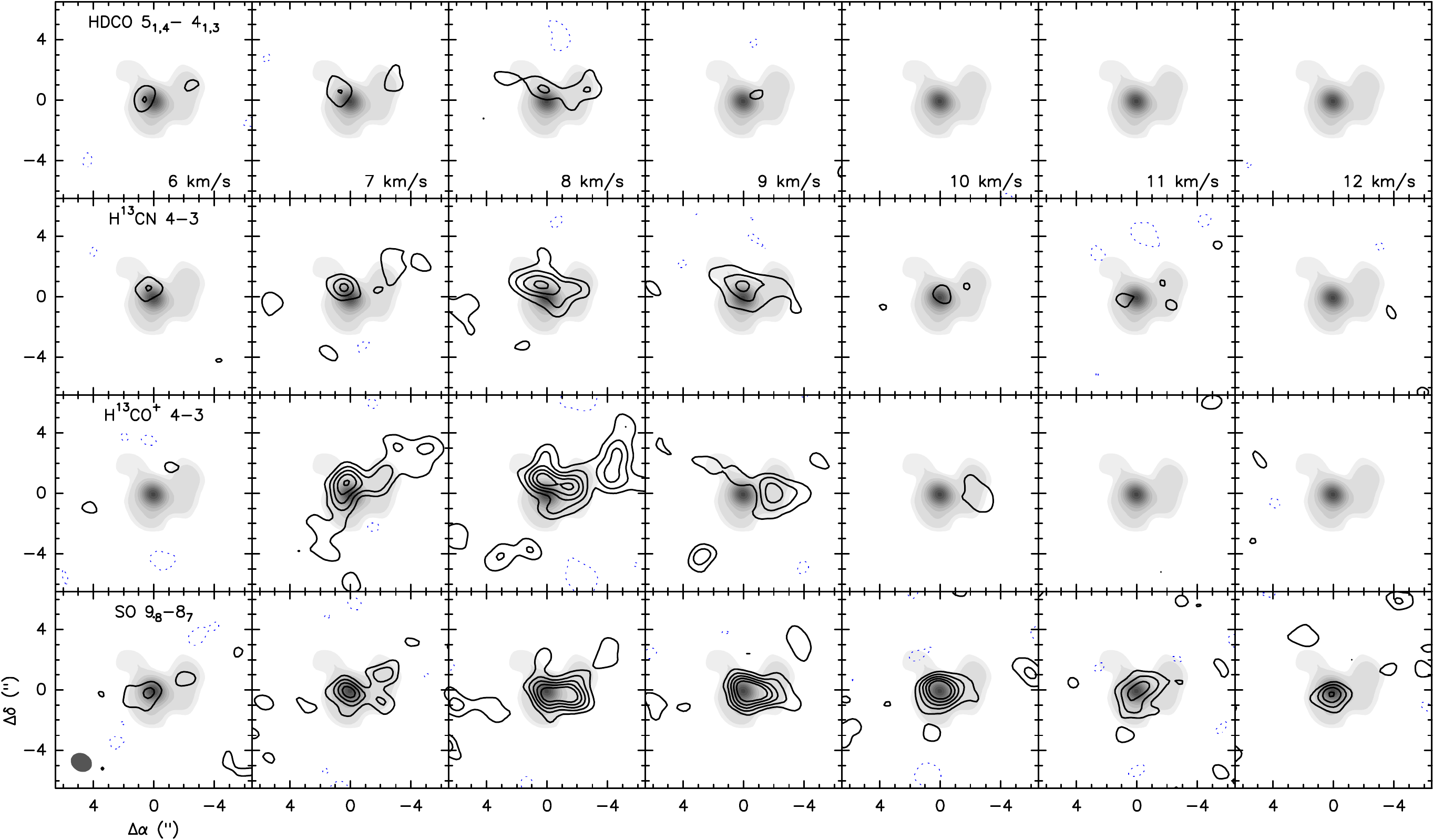}
\caption{
Channel velocity contour maps of HDCO (5$_{1, 4}$--4$_{1, 3}$),  \htcn~(4--3), \htcop~(4--3) and SO (9$_8$--8$_7$) from the SMA observations. The contour levels are $-3$,\,3,\,5,\,7,\,9\,11 times the rms noise of the maps, 0.19\,\jybm. The SMA 880\,$\micron$ dust emission map at an angular resolution of $1\farcs2$ is also shown in  the gray scale. 
}
\label{fig:Fmol}
\end{figure*}

Table~\ref{table:T1} shows the list of the molecular lines detected by the SMA toward SMM1, including a number of dense molecular tracers.  Figure \ref{fig:Fmol} shows the channel maps of the molecules tracing the dense molecular core.  The HDCO, \htcn and \htcop lines trace mostly the region north of the SMM1 peak. The emission peaks at $\sim$\,8\,\kms, which is slightly lower than the $\sim$\,8.5\,\kms velocity of the clump surrounding the cores \citep[][]{Lee2014}. The dust peak appears to be mostly devoid of emission from these three lines; this has also been observed in other cores, which are usually hot or warm \citep[e.g.,][]{Rao2009, Girart2013}.   The emission is mainly detected only in the 7--9\,\kms velocity range, suggesting that the gas is relatively quiescent.  In contrast, the SO emission appears to have a significantly broader emission, spanning over 5\,\kms, and being brighter at the dust emission peak of SMM1-a. This suggests that SO is a good tracer of the warmer and denser molecular environment around SMM1-a or, alternatively, that it has been excited by shocks in the outflow.

%%%%%%%%%%% TABLE %%%%%%%%%%%%
\begin{table}[hbt!]
\begin{center}
\normalsize
\caption{\normalsize \vspace{0.1in} Molecular lines detected by the SMA}
\label{table:T1}
\begin{tabular}{lcc}
\hline
Molecular 				& $\nu$			& E$_l$	 \\
transition				& (GHz)			& (K)	 \\
\hline
HDCO  5$_{1, 4}$--4$_{1, 3}$ 
						& 335.09678 	& 40.17    \\
HC$^{15}$N (4--3)$^a$	& 344.20011 	& 24.78    \\
H$^{13}$CN (4--3)		& 345.33976	& 24.86    \\
CO (3--2)					& 345.79599	& 16.60    \\
SO  (9$_8$--8$_7$)		& 346.52848	& 62.14    \\
H$^{13}$CO$^+$ (4--3)	& 346.99835	& 24.98    \\
SiO (8--7)				& 347.33082	& 58.35    \\
\hline
\end{tabular}
\\
\vspace{0.05in}
$^a$  Observed only in the compact configuration on 2012 May 25.
\end{center}
\end{table}
%%%%%%%%%%%%%%%%%%%%%%%%%%%%

\end{document}